\documentclass[runningheads]{llncs}
\usepackage{graphicx}
\usepackage{amsmath}
\usepackage{tikz}
\usepackage{pgfplots}
\pgfplotsset{width=7cm,compat=1.3}
\usepackage{float}
\usepackage{wrapfig}
\usepackage[inline]{enumitem}

\usepackage{hyperref}
\usepackage{cite}

\hypersetup{
    colorlinks=true,
    linkcolor=blue,
    filecolor=magenta,      
    urlcolor=blue,
}

\usepackage{xcolor,colortbl}
\definecolor{Gray}{gray}{0.85}
\definecolor{LightCyan}{rgb}{0.88,1,1}

\newcolumntype{a}{>{\columncolor{Gray}}c}
\newcolumntype{b}{>{\columncolor{white}}c}

\begin{document}
\title{CellCentroidFormer: Combining Self-attention and Convolution for Cell Detection}
\titlerunning{CellCentroidFormer}
%
%
\author{Royden Wagner \and
Karl Rohr}
\authorrunning{Wagner and Rohr}
%
\institute{
Biomedical Computer Vision Group, BioQuant, IPMB, Heidelberg University, Germany\\
\email{royden.wagner@bioquant.uni-heidelberg.de} \newline \email{k.rohr@uni-heidelberg.de}}
\maketitle              
\begin{abstract}
Cell detection in microscopy images is important to study how cells move and interact with their environment. 
Most recent deep learning-based methods for cell detection use convolutional neural networks (CNNs).
However, inspired by the success in other computer vision applications, vision transformers (ViTs) are also used for this purpose.
We propose a novel hybrid CNN-ViT model for cell detection in  microscopy images to exploit the advantages of both types of deep learning models. 
We employ an efficient CNN, that was pre-trained on the \mbox{ImageNet} dataset, to extract image features and utilize transfer learning to reduce the amount of required training data. 
Extracted image features are further processed by a combination of convolutional and transformer layers, so that the convolutional layers can focus on local information and the transformer layers on global information.
Our centroid-based cell detection method represents cells as ellipses and is end-to-end trainable.
Furthermore, we show that our proposed model can outperform fully convolutional one-stage detectors on four different 2D microscopy datasets. 
Code is available at: \newline \url{https://github.com/roydenwa/cell-centroid-former}

\end{abstract}
{
\setlength{\belowcaptionskip}{-15pt}
\begin{figure}[H]
    \centering
    \resizebox{\columnwidth}{!} {
    \includegraphics{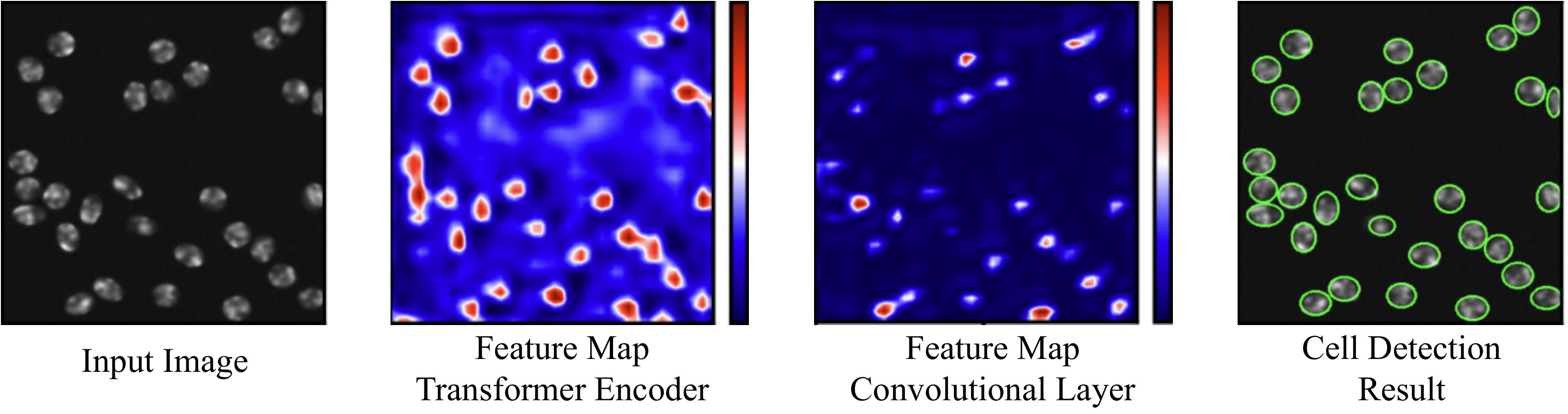}
    }
    \caption{\textbf{Feature maps and cell detection results for a fluorescence microscopy image.} The feature map on the left was generated by a transformer encoder in the neck part of our model, the feature map on the right was generated  by an adjacent convolutional layer. The transformer encoder focuses on the overall cell shapes (global features), while the convolutional layer focuses on the cell centroids (local features). Both feature maps have been enlarged, their original size is $48 \times 48$ pixels.}
    \label{fig:featuremaps}
\end{figure}
}
\section{Introduction}
Cell detection is an important task when studying biomedical microscopy images and time-lapse microscopy videos.
Main applications are the quantification of cellular structures as well as studying how cells move and interact with their environment.
Most recent deep learning-based methods for cell detection in microscopy images use convolutional neural networks (CNNs) (e.g., \cite{hung2020keras}, \cite{jiang2020geometry}, \cite{nishimura2021semi}).
However, inspired by the success in other computer vision applications, vision transformers (ViTs) (e.g., \cite{prangemeier2020}) are also used for this purpose. 

The comparison of ViTs and CNNs in computer vision applications reveals that the receptive fields of ViTs and CNNs are fundamentally different \cite{raghu2021}.
The receptive fields of ViTs capture local and global information in both earlier and later layers.
The receptive fields of CNNs, on the other hand, initially capture local information and gradually grow to capture global information in later layers.
This suggests that ViTs better preserve spatial information across the generated feature maps within the models, which is advantageous for object detection.
However, a limitation of most current ViT models is that they require much more training data to reach or surpass the performance of CNNs in computer vision tasks \cite{khan2021}.
This is a major limitation for biomedical applications, where annotated training samples are limited.
Another limitation of ViTs is that their core mechanism, multi-head self-attention, has a computational complexity of $O(n^2)$, where $n$ is the length of the input sequence.
Consequently, large computational resources are required for training such models.

To exploit the advantages of both types of deep learning models, we propose a hybrid CNN-ViT model for cell detection in microscopy images. 
We employ an efficient CNN, that was pre-trained on the ImageNet \cite{deng2009} dataset, to extract images features and utilize transfer learning to reduce amount of required training data. 
Extracted image features are further processed by a combination of convolutional and transformer layers, so that the convolutional layers can focus on local information and the transformer layers on global information.
We propose a one-stage cell detection method that is end-to-end trainable.
Overall, the contributions of our paper are twofold:
\begin{enumerate}
    \item We introduce a novel deep learning model that combines self-attention and convolution for cell detection in microscopy images.
    \item We propose a centroid-based cell detection method that represents cells as ellipses.
\end{enumerate}

\section{Related Work}
Several related methods use CNNs \cite{xie2018microscopy, jiang2020geometry, nishimura2021semi, wollmann2021, fujii2021cell, hung2020keras, li2021detection} or combine CNNs with classical image analysis approaches \cite{tyson2021deep} for cell detection in biomedical microscopy images.
Amongst these methods, two approaches are common: \begin{enumerate*}[label=(\roman*)]
\item to perform cell detection by predicting a heatmap for cell positions \cite{xie2018microscopy, nishimura2021semi, fujii2021cell} or
\item to perform cell detection by predicting the coordinates of bounding boxes for cells \cite{hung2020keras, jiang2020geometry, li2021detection}.
\end{enumerate*}
The heatmap-based methods are well suited for cell counting, but are unsuitable for morphological studies since they only predict where cells are located and do not provide information on the cell dimensions.
The bounding box-based methods are more flexible as they predict the cell positions and the cell dimensions (width and height).
However, the bounding box-based methods use a cascade of two CNNs and are therefore not end-to-end trainable.
To deal with sparsely annotated or small datasets, semi-supervised learning \cite{nishimura2021semi, fujii2021cell} or pre-training with synthetic data \cite{xie2018microscopy} is used. 
Thus complicating the training process, since pseudo-labels must be iteratively generated or synthetic data must be created in advance.

The Cell-DETR \cite{prangemeier2020} model is architecturally most related to ours. 
Cell-DETR is a hybrid CNN-ViT model for cell detection and segmentation.
Prangemeier et al. use a CNN backbone to extract image features, a transformer encoder-decoder block to process image features and a model head with a multi-layer perceptron for cell detection.
Due to the high computational complexity of the transformer encoder-decoder block, they use a small input size of $128\times128$ pixels.
Therefore, a considerable amount of information is lost when downsizing high-resolution microscopy images.

\section{Method}
\subsection{Model Architecture}
The proposed hybrid CNN-ViT model combines self-attention and convolution for cell detection.
For this purpose, we use MobileViT blocks \cite{mehta2022mobilevit}, which combine transformer encoders \cite{dosovitskiy2021an, vaswani2017attention} with convolutional layers.
In MobileViT blocks, input tensors are first processed by convolutional layers, then, the extracted features are unfolded into a sequence and passed through a transformer encoder.
Finally, the output tensor of the transformer encoder is folded again into a 3D representation and concatenated with the input tensor.
Thereby, the convolutional layers extract local information and the self-attention mechanism in the transformer encoder associates local information from distant features to capture global information.
We use MobileViT blocks in the neck part of our proposed model to enhance global information compared to a fully convolutional neck part.
MobileViT blocks are a light-weight alternative to the original transformer encoder-decoder design \cite{vaswani2017attention}.
However, due to their multi-head self-attention layers, MobileViT blocks still have a much higher computational complexity (CC) than convolutional layers:
\begin{equation}
CC_{mhs-attn} \cong \begin{cases}
    O(n^2)         & \text{if } n >> d \cdot h,\\
    O(n^2 \cdot d \cdot h)    & \text{else}
\end{cases}
\end{equation}
\begin{equation}
CC_{conv} \cong \begin{cases}
    O(n)         & \text{if } n >> d \cdot k \cdot f,\\
    O(n \cdot d \cdot k \cdot f)    & \text{else}
\end{cases}
\end{equation}
where $n$ is the sequence length, $d$ is the sequence depth, $h$ is the number of self-attention heads, $k$ the size of the convolutional kernels, and $f$ is the number of convolutional filters.
Thus, we combine MobileViT blocks in the neck part of our model with convolutional layers to extract more features without increasing the computational complexity excessively. 
In addition, we add layer normalization layers for regularization and to allow higher learning rates during training.

\begin{figure}[H]
    \centering
    \resizebox{\columnwidth}{!} {
    \includegraphics{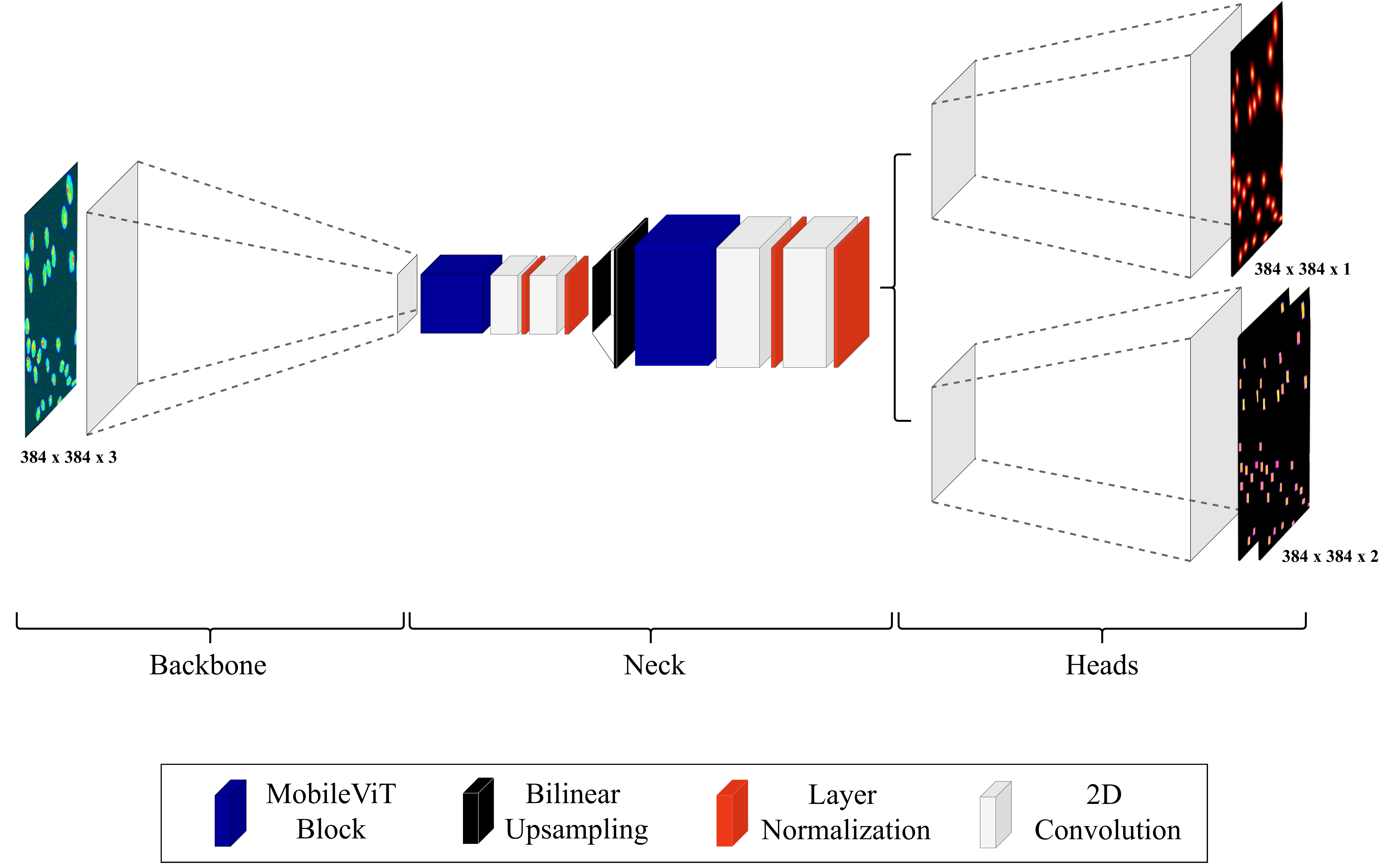}
    }
    \caption{\textbf{CellCentroidFormer model.} \textbf{Backbone:} Five blocks of an EfficientNetV2S. \textbf{Neck:} MobileViT blocks and convolutional layers. \textbf{Heads:} Fully convolutional upsampling blocks.}
    \label{fig:model}
\end{figure}

As backbone of our proposed model, we use parts of an EfficientNetV2S \cite{tan2021efficientnetv2} CNN model.
EffcientNet models consist of six high-level blocks, we use five of these blocks to extract image features.
We initialize the backbone with weights learned from training on the ImageNet dataset to leverage transfer learning and reduce the amount of required training data.
EfficientNetV2S models are optimized for a fixed input size of $384\times384\times3$ pixels.
Therefore, we resize all input images to this input size.
We represent cells by their centroid, their width, and their height. 
Our model contains two fully convolutional heads to predict these cell properties.
The heads contain 2D convolution, batch normalization, and bilinear upsampling layers.
We do not use further MobileViT blocks in the heads to reduce the computational complexity of our model, and since later convolutional layers have a large receptive field that allows them to capture global information \cite{raghu2021}.
The first head predicts a heatmap for cell centroids, and the second head predicts the cell dimensions (width and height) at the position of the corresponding cell centroid.
The output dimensions of our model are $384\times384$, thus, the output stride is one and we do not need an additional offset head to account for offset errors (as in, e.g., \cite{yang2020circlenet}).
Figure \ref{fig:model} shows the overall model architecture.
\subsection{Bounding Ellipses for Cell Detection}
\begin{figure}[h]
    \centering
    \resizebox{\columnwidth}{!} {
    \includegraphics{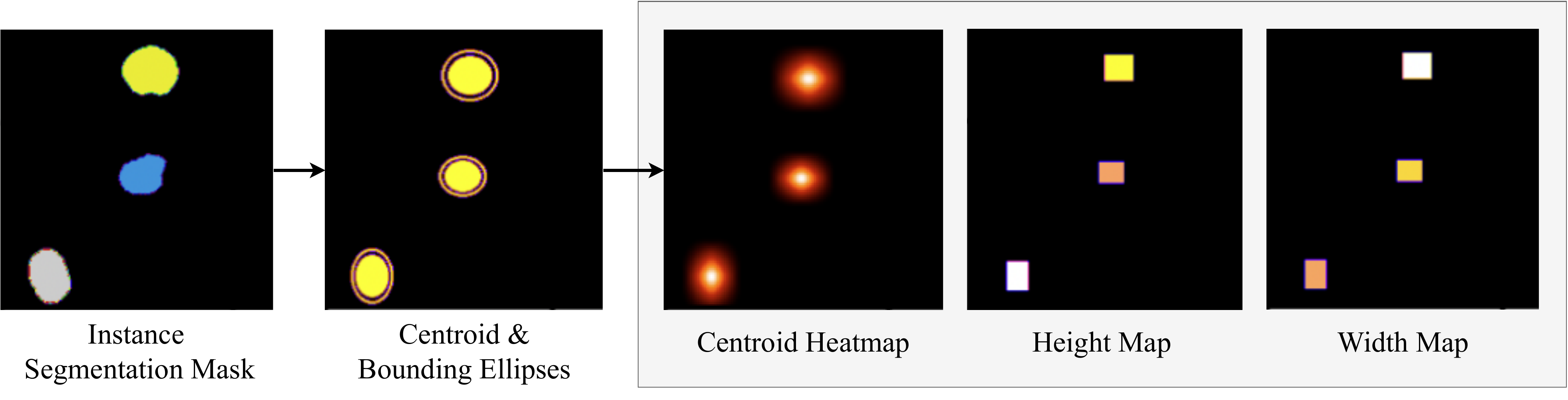}
    }
    \caption{\textbf{Centroid representation of bounding ellipses.} Instance segmentation masks are converted into training samples for cell detection. The training samples contain centroid heatmaps, height maps, and width maps.}
    \label{fig:trainsamples}
\end{figure}

Yang et al. \cite{yang2020circlenet} argue that traditional bounding box-based object detection methods are not optimized for biomedical applications. 
They state that most relevant objects in biomedical applications have round shapes, and thus they propose using bounding circles instead of bounding boxes.
We extend this idea and use \textit{bounding ellipses} for cell detection, since most cell types in biomedical applications have an ellipsoidal shape.
Figure \ref{fig:trainsamples} shows how we generate training samples for the proposed centroid-based cell detection method.
We approximate the centroid of a cell by an ellipse with width and height adjusted to the corresponding cell dimensions. 
We blur the centroid ellipses using a Gaussian kernel to reduce the loss value when a centroid is only missed by a small distance.
Similar as in \cite{zhou2019objects}, the cell width and cell height are encoded in rectangles located at the position of the cell centroids.

\section{Experiments}
\subsection{Datasets and Pre-processing}
\begin{figure}[h]
    \centering
    \resizebox{\columnwidth}{!} {
    \includegraphics{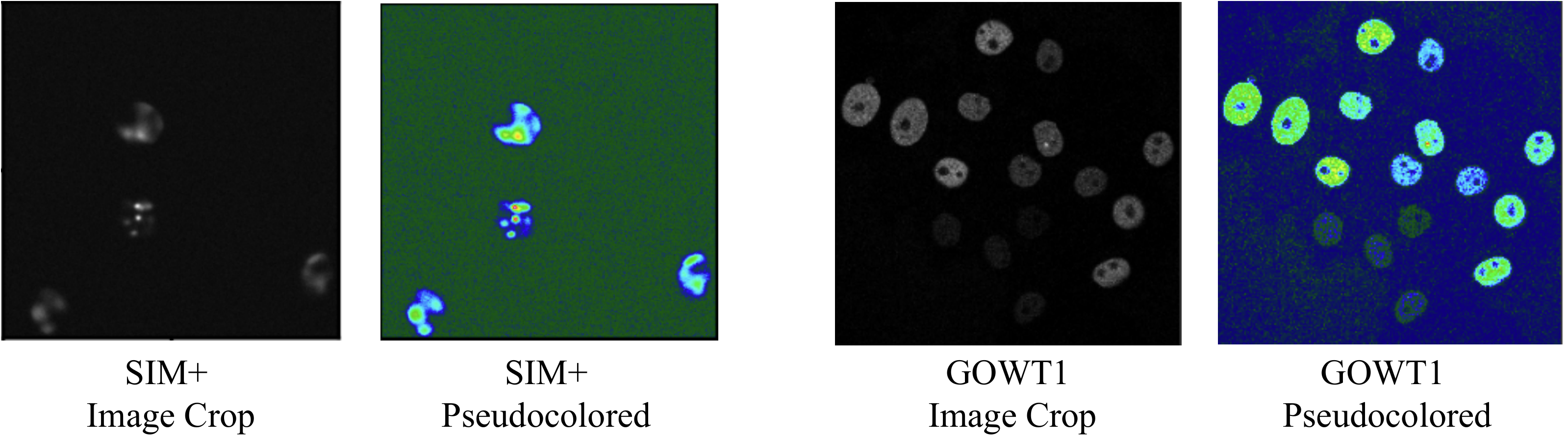}
    }
    \caption{\textbf{Pseudocolored image crops of two considered datasets.}}
    \label{fig:trainsamples2}
\end{figure}

We evaluate our hybrid CNN-ViT model using four different 2D microscopy datasets from the Cell Tracking Challenge \cite{ulman2017objective}. 
The Fluo-N2DL-HeLa (HeLa) dataset consists of 2D fluorescence microscopy images of cells stably expressing H2b-GFP.
The images of this dataset have a size of $700 \times 1100$ pixels.
The Fluo-N2DH-SIM+ (SIM+) dataset consists of simulated 2D fluorescence microscopy images of HL60 cells stained with Hoechst dyes. 
The images of this dataset have a size of $773\times739$ pixels.
The Fluo-N2DH-GOWT1 (GOWT1) dataset consists of 2D fluorescence microscopy images of mouse stem cells.
These images have a size of $1024\times1024$ pixels.
The PhC-C2DH-U373 (U373) dataset consists of 2D phase contrast microscopy images of glioblastoma-astrocytoma U373 cells on a polyacrylamide substrate.
The images of this dataset have a size of $520 \times 696$ pixels.
Similar as in \cite{wagner2022}, we use pseudocoloring to generate input images with three channels suitable for the pretrained backbone.
We use the ncar pseudo spectral colormap\footnote{\scriptsize\url{https://www.ncl.ucar.edu/Document/Graphics/ColorTables/MPL\_gist\_ncar.shtml}}, which is well suited for coloring cell nuclei and their immediate surroundings to distinguish these regions from the background.
Figure \ref{fig:trainsamples2} shows pseudocolored image crops of the SIM+ and GOWT1 datasets.

\subsection{Baseline Model}
Similar to semantic segmentation, our method treats cell detection as a pixel-wise regression objective.
Therefore, we choose a CNN designed for semantic segmentation as baseline model in the following experiments.
The Dual U-Net model \cite{li2019dual} has an encoder-decoder architecture with two decoders.
Most recently, the Dual U-Net model was used as part of a tracking and segmentation method \cite{scherr2021improving} that achieved multiple top-3 rankings in the Cell Tracking Challenge\footnote{\scriptsize\url{http://celltrackingchallenge.net/participants/KIT-Sch-GE}}.
Analogously to our model, we use one decoder of the Dual U-Net model to predict the centroid heatmap and the other to predict the cell dimensions.

\subsection{Training Setup, Metrics, and Hyperparameters}
We use geometric data augmentations such as grid distortion, elastic transformation, shifting, and rotation to increase the size of each dataset to 2150 samples.
The resulting datasets are each split into a training dataset (80\%), a validation dataset (10\%), and a test dataset (10\%).
All images are normalized using min-max scaling.
Our CellCentroidFormer model is trained with pseudocolored training samples, the Dual U-Net model with normalized grayscale training samples.
We use three Huber loss functions \cite{huber1964robust}, one loss function per output, to train both models. 
The total loss is computed by a weighted sum of the three loss values:
\begin{equation}
\mathcal{L}_{Huber}(y, \widehat{y}) = \begin{cases}
    \frac{1}{2}(y - \widehat{y})^2        & \text{if } y - \widehat{y} \leq 1.0,\\
    (y - \widehat{y}) - \frac{1}{2}    & \text{else}
\end{cases}
\end{equation}
\begin{equation}
\mathcal{L}_{total} = \mathcal{L}_{heatmap} + \frac{1}{2} \cdot \mathcal{L}_{height} + \frac{1}{2} \cdot \mathcal{L}_{width}
\end{equation}
The centroid heatmap loss ($\mathcal{L}_{heatmap}$) contributes the most to the total loss because our method is inherently centroid-based.
When decoding the predictions of our model, the width and height of the cells are only looked up at positions where cell centroids are located according to the centroid heatmap prediction.

As performance metrics, we use the mean intersection over union (MeanIoU) and the structural similarity metric (SSIM) \cite{wang2004image}.
We convert the centroid heatmaps to binary masks to evaluate the detection performance.
Therefore, we apply thresholding to the heatmaps. 
Afterwards, we compute the MeanIoU value of the predicted and the ground truth centroid mask:
\begin{equation}
MeanIoU = \frac{1}{C} \sum_{C} \frac{TP_C}{TP_C + FP_C + FN_C}
\end{equation}
For the two class labels (C), background and cell centroids, the true positive (TP), false positive (FP), and false negative (FN) pixels are determined.

For the cell dimensions (height and width), we use the SSIM metric to quantify the performance. 
The metric measures the similarity between two images or matrices and is defined as:
\begin{equation}
SSIM(x_1, x_2) = \frac{(2\mu_{x_1}\mu_{x_2} + (0.01L)^2)(2\sigma_{x_1x_2} + (0.03L)^2)}
{(\mu_{x_1}^2 + \mu_{x_2}^2 + (0.01L)^2)(\sigma_{x_1}^2 + \sigma_{x_2}^2 + (0.03L)^2)}
\end{equation}
For the two inputs ($x_1$ and $x_2$), the mean ($\mu$), the variance ($\sigma$), and the dynamic range ($L$) are computed.

We train both models per dataset for 50 epochs with a batch size of 4 using a Nvidia{\tiny\textregistered} V100 GPU.
We use Adam \cite{kingma2014adam} as optimizer with an initial learning rate of $1^{-4}$ and reduce the learning rate at plateaus.
\subsection{Training Curves and Performance Comparison}
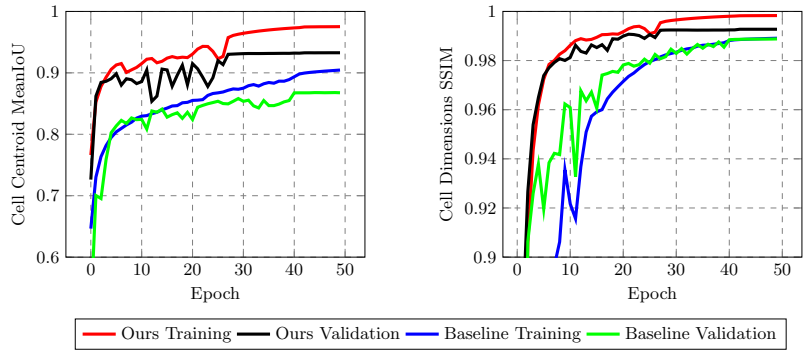
\begin{figure}[H]
\pgfplotstableread[col sep=comma,]{assets/ccf_hela_train_dynamics.csv}\ccf
\pgfplotstableread[col sep=comma,]{assets/dunet_hela_train_dynamics.csv}\dunet
{\resizebox{0.9\columnwidth}{!} {
    \hspace{0.5cm}
    \begin{tikzpicture}
    \pgfplotsset{grid style={dashed,gray}}
    \begin{axis}[
      ymin=0.6,ymax=1,
      name=ax1,
      ymajorgrids=true,
      xmajorgrids=true,
      xlabel={Epoch},
      xlabel near ticks,
      every axis plot/.append style={ultra thick},
      ylabel={Cell Centroid MeanIoU},
       legend style={font=\small, at={(0.03, -0.25)},anchor=north west,legend columns=4}
    ]
      \addplot[red] table [x=epoch, y=centroid_heatmap_mean_iou_thresh]{\ccf};
      \addlegendentry{Ours Training};
      \addplot[black] table [x=epoch, y=val_centroid_heatmap_mean_iou_thresh]{\ccf};
      \addlegendentry{Ours Validation};
      \addplot[blue] table [x=epoch, y=centroid_heatmap_mean_iou_thresh]{\dunet};
      \addlegendentry{Baseline Training};
      \addplot[green] table [x=epoch, y=val_centroid_heatmap_mean_iou_thresh]{\dunet};
      \addlegendentry{Baseline Validation};
    \end{axis}
    
    \hspace{2.5cm}
    
    \begin{axis}[
     at={(ax1.south east)},
      ymin=0.9,ymax=1,
      ymajorgrids=true,
      xmajorgrids=true,
      xlabel={Epoch},
      xlabel near ticks,
      ylabel={Cell Dimensions SSIM},
      ylabel near ticks,
      every axis plot/.append style={ultra thick}
    ]
    \addplot[red] table [x=epoch, y=dimensions_ssim]{\ccf};
    \addplot[black] table [x=epoch, y=val_dimensions_ssim]{\ccf};
    \addplot[blue] table [x=epoch, y=dimensions_ssim]{\dunet};
    \addplot[green] table [x=epoch, y=val_dimensions_ssim]{\dunet};
    \end{axis}
    \end{tikzpicture}
}} 
\caption{\textbf{Training curves for the HeLa dataset.}}
\label{fig:traincurves}
\end{figure}

Figure \ref{fig:traincurves} shows the training curves for the HeLa dataset.
Our model converges around the 30th epoch, whereas the baseline model converges around the 40th epoch.
In these training epochs, the lowest learning rate of $1^{-6}$ is also reached for both models and, accordingly, the metrics do not change much in the following epochs.
Our model converges faster since the pretrained backbone was already trained to extract image features.
The performance difference for the cell centroid MeanIoU score is greater than for the cell dimensions SSIM score, but overall our model yields higher values for all considered metrics on both datasets (training and validation).

\begin{table}[h!]
\caption{\textbf{Performance on different microscopy datasets.} Metrics are evaluated after training for 50 epochs with the corresponding training datasets.
}
\centering
\resizebox{\columnwidth}{!} {
\begin{tabular}{l | a | b | a | b }
\hline
\rowcolor{white}
Dataset & Model  & Backbone & Centroid MeanIoU $\Uparrow$ & Dimensions SSIM $\Uparrow$ \\
\hline
SIM+ Test & Dual U-Net & - & 0.8033 & 0.9631 \\
                         & CircleNet & EfficientNetV2S & 0.8308 & 0.9011  \\
                         & CellCentroidFormer & EfficientNetV2S & \textbf{0.8492} & \textbf{0.9858} \\
                         \hline
GOWT1 Test & Dual U-Net & - & 0.9278 & 0.9909 \\
                         & CircleNet & EfficientNetV2S & 0.9108 & 0.9192 \\
                         & CellCentroidFormer & EfficientNetV2S & \textbf{0.9355} & \textbf{0.9959}  \\
                         \hline
U373 Test & Dual U-Net & - & 0.9170 & 0.9908 \\
                         & CircleNet & EfficientNetV2S & 0.8802 & 0.9241 \\
                         & CellCentroidFormer & EfficientNetV2S & \textbf{0.9256} & \textbf{0.9923} \\
                         \hline
HeLa Test & Dual U-Net & - & 0.8675 & 0.9885 \\
                         & CircleNet & EfficientNetV2S & 0.7650 & 0.7507 \\
                         & CellCentroidFormer & EfficientNetV2S & \textbf{0.9287} & \textbf{0.9937} \\
                         \hline
\end{tabular}
}
\label{table:comparison}
\end{table}

Table \ref{table:comparison} shows a comparison the performance of the two models on the used \textit{test} datasets.
Additionally, we train a CircleNet \cite{yang2020circlenet} model with an EfficientNetV2S as backbone.
As in \cite{xiao2018simple, zhou2019objects}, we combine the backbone with upsampling blocks, such that the CircleNet has an output stride of 4.
CircleNets detect bounding circles for cells by predicting a heatmap for cell centers, the circle radius, and a local offset.
We train the model with pseudocolored samples, use a Huber loss for the center heatmap predictions, and keep the rest of the training setup as in \cite{yang2020circlenet}.
To compute the SSIM score for the cell dimensions, we compute the SSIM score for the radius prediction and cell width map, the SSIM score for the radius prediction and cell height map, and average them.

\begin{figure}
    \centering
    \resizebox{\columnwidth}{!} {
    \includegraphics{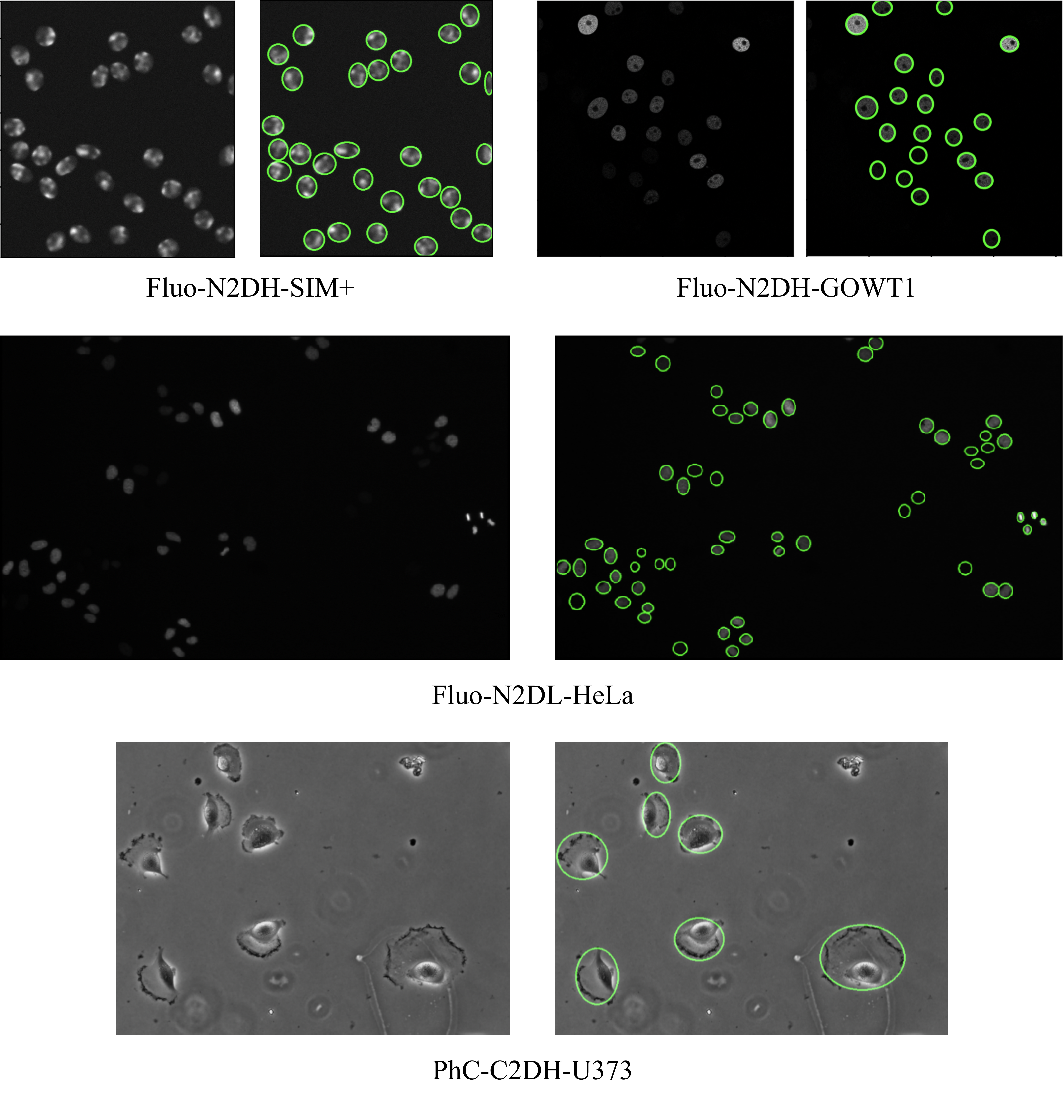}
    }
    \caption{\textbf{Example microscopy images and cell detection results of our proposed method.}}
    \label{fig:detresults}
\end{figure}

Our CellCentroidFormer model outperforms the Dual U-Net and the \mbox{CircleNet} on all considered datasets.
As in Figure \ref{fig:traincurves}, the performance difference for the cell centroid MeanIoU score is greater than for the cell dimensions SSIM score.
Our model and the Dual U-Net model yield higher cell dimensions SSIM scores than the CircleNet model since most cells have an ellipsoidal shape, which is more accurately represented by an ellipse than by a circle.
On the SIM+ dataset, our model and the CircleNet model outperform the Dual U-Net model on the cell centroid MeanIoU score.
On the HeLa dataset, the CircleNet performs worst because this dataset contains many small cells (see Figure \ref{fig:detresults}), that are challenging to distinguish in a low resolution output of 128 $\times$ 128.

Table \ref{table:times} shows a comparison of the training times, the inference times, and the overall number of parameters per model.
As previously shown \cite{mehta2022mobilevit}, vision \mbox{transformers} need more time for training and inference than CNNs. 
\begin{wraptable}[9]{r}{0.5\columnwidth}
\vspace{-25pt}
\caption{\textbf{Model comparison.} Inference times ($TI_{GPU}$) are measured end-to-end and include the data transfer between the host (CPU) and the \mbox{device (GPU)}.\vspace{9pt}}
\resizebox{0.5\columnwidth}{!} {
\begin{tabular}{l | a | b | a}
\rowcolor{white}
\hline
Model & T$_{Epoch}$ $\Downarrow$ & TI$_{GPU}$ $\Downarrow$ & \#Params $\Downarrow$\\
\hline
Dual U-Net & 106\,s & 67\,ms & 33.0\,M  \\
CircleNet & 64\,s & 59\,ms & 24.6\,M \\
CellCentroidFormer & 122\,s & 82\,ms & 11.5\,M \\
\hline
\end{tabular}
}
\label{table:times}
\end{wraptable}
This also applies to our hybrid model, thus, it requires more time to train our model for one epoch (T$_{Epoch}$), and the inference time for one input image is somewhat slower (TI$_{GPU}$).
However, our model only requires one third of the parameters of the Dual U-Net model and roughly half of the parameters of the CircleNet model (11.5\,M vs. 33\,M vs. 24.6\,M) to achieve superior performance.

Figure \ref{fig:detresults} shows example images of the considered datasets and the corresponding cell detection results of our model.

\section{Conclusion}
We have presented a novel hybrid CNN-ViT model that combines self-attention and convolution for cell detection.
Our centroid-based cell detection method represents cells as ellipses and is end-to-end trainable.
We have shown that the transformer layers in the neck part of our proposed model can focus on the overall cell shapes, while adjacent convolutional layers focus on the cell centroids.
Our experiments reveal that pseudocoloring in combination with pretrained backbones improves the cell detection performance, whereas larger output strides worsen the performance.
Furthermore, our CellCentroidFormer model outperforms fully convolutional one-stage detectors on four different 2D microscopy datasets despite having less parameters overall.
\subsubsection*{Acknowledgements.}
Support of the DFG (German Research Foundation) within the SFB 1129 (project Z4) and the SPP 2202 (RO 2471/10-1), and the BMBF (German Federal Ministry of Education and Research) within the de.NBI is gratefully acknowledged.
\newpage
\bibliographystyle{splncs04}
\bibliography{references}

\begin{thebibliography}{10}
\providecommand{\url}[1]{\texttt{#1}}
\providecommand{\urlprefix}{URL }
\providecommand{\doi}[1]{https://doi.org/#1}

\bibitem{deng2009}
Deng, J., Dong, W., Socher, R., Li, L.J., Li, K., Fei-Fei, L.: {ImageNet: A
  large-scale hierarchical image database}. In: Conference on Computer Vision
  and Pattern Recognition (CVPR). pp. 248--255. IEEE (2009)

\bibitem{dosovitskiy2021an}
Dosovitskiy, A., Beyer, L., Kolesnikov, A., Weissenborn, D., Zhai, X.,
  Unterthiner, T., Dehghani, M., Minderer, M., Heigold, G., Gelly, S.,
  Uszkoreit, J., Houlsby, N.: {An Image is Worth 16x16 Words: Transformers for
  Image Recognition at Scale}. In: International Conference on Learning
  Representations (ICLR) (2021)

\bibitem{fujii2021cell}
Fujii, K., Suehiro, D., Nishimura, K., Bise, R.: {Cell Detection from Imperfect
  Annotation by Pseudo Label Selection Using P-classification}. In:
  International Conference on Medical Image Computing and Computer-Assisted
  Intervention \mbox{(MICCAI)}. pp. 425--434. Springer (2021)

\bibitem{huber1964robust}
Huber, P.J.: {Robust Estimation of a Location Parameter}. Annals of Statistics
  \textbf{53},  73--101 (1964)

\bibitem{hung2020keras}
Hung, J., Goodman, A., Ravel, D., Lopes, S.C., Rangel, G.W., Nery, O.A.,
  Malleret, B., Nosten, F., Lacerda, M.V., Ferreira, M.U., et~al.: Keras r-cnn:
  library for cell detection in biological images using deep neural networks.
  BMC bioinformatics  \textbf{21}(1), ~1--7 (2020)

\bibitem{jiang2020geometry}
Jiang, H., Li, S., Liu, W., Zheng, H., Liu, J., Zhang, Y.: {Geometry-Aware Cell
  Detection with Deep Learning}. Msystems  \textbf{5}(1),  e00840--19 (2020)

\bibitem{khan2021}
Khan, S., Naseer, M., Hayat, M., Zamir, S.W., Khan, F.S., Shah, M.:
  {Transformers in Vision: A Survey}. ACM Comput. Surv.  (2021)

\bibitem{kingma2014adam}
Kingma, D.P., Ba, J.: {Adam: A Method for Stochastic Optimization}.
  International Conference on Learning Representations (ICLR)  (2015)

\bibitem{li2021detection}
Li, X., Xu, Z., Shen, X., Zhou, Y., Xiao, B., Li, T.Q.: {Detection of Cervical
  Cancer Cells in Whole Slide Images Using Deformable and Global Context Aware
  Faster RCNN-FPN}. Current Oncology  \textbf{28}(5),  3585--3601 (2021)

\bibitem{li2019dual}
Li, X., Wang, Y., Tang, Q., Fan, Z., Yu, J.: {Dual U-Net for the Segmentation
  of Overlapping Glioma Nuclei}. IEEE Access  \textbf{7},  84040--84052 (2019)

\bibitem{mehta2022mobilevit}
Mehta, S., Rastegari, M.: {MobileViT: Light-weight, General-purpose, and
  Mobile-friendly Vision Transformer}. In: International Conference on Learning
  Representations (ICLR) (2022)

\bibitem{nishimura2021semi}
Nishimura, K., Cho, H., Bise, R.: {Semi-supervised Cell Detection in Time-Lapse
  Images Using Temporal Consistency}. In: International Conference on Medical
  Image Computing and Computer-Assisted Intervention (MICCAI). pp. 373--383.
  Springer (2021)

\bibitem{prangemeier2020}
Prangemeier, T., Reich, C., Koeppl, H.: {Attention-Based Transformers for
  Instance Segmentation of Cells in Microstructures}. In: International
  Conference on Bioinformatics and Biomedicine (BIBM). IEEE (2020)

\bibitem{raghu2021}
Raghu, M., Unterthiner, T., Kornblith, S., Zhang, C., Dosovitskiy, A.: {Do
  Vision Transformers See Like Convolutional Neural Networks?} In: Advances in
  Neural Information Processing Systems (NeurIPS) (2021)

\bibitem{scherr2021improving}
Scherr, T., L{\"o}ffler, K., Neumann, O., Mikut, R.: {On Improving an Already
  Competitive Segmentation Algorithm for the Cell Tracking Challenge-Lessons
  Learned}. bioRxiv  (2021)

\bibitem{tan2021efficientnetv2}
Tan, M., Le, Q.: {EfficientNetV2: Smaller models and faster training}. In:
  International Conference on Machine Learning (ICML). pp. 10096--10106. PMLR
  (2021)

\bibitem{tyson2021deep}
Tyson, A.L., Rousseau, C.V., Niedworok, C.J., Keshavarzi, S., Tsitoura, C.,
  Cossell, L., Strom, M., Margrie, T.W.: {A deep learning algorithm for 3D cell
  detection in whole mouse brain image datasets}. PLoS Computational Biology
  \textbf{17}(5),  e1009074 (2021)

\bibitem{ulman2017objective}
Ulman, V., Ma{\v{s}}ka, M., Magnusson, K.E., Ronneberger, O., Haubold, C.,
  Harder, N., Matula, P., Matula, P., Svoboda, D., Radojevic, M., et~al.: An
  objective comparison of cell-tracking algorithms. Nature Methods
  \textbf{14}(12),  1141--1152 (2017)

\bibitem{vaswani2017attention}
Vaswani, A., Shazeer, N., Parmar, N., Uszkoreit, J., Jones, L., Gomez, A.N.,
  Kaiser, {\L}., Polosukhin, I.: Attention is all you need. Advances in neural
  information processing systems (NeurIPS)  \textbf{30} (2017)

\bibitem{wagner2022}
Wagner, R., Rohr, K.: {EfficientCellSeg: Efficient Volumetric Cell Segmentation
  Using Context Aware Pseudocoloring}. In: Medical Imaging with Deep Learning
  (MIDL) (2022)

\bibitem{wang2004image}
Wang, Z., Bovik, A.C., Sheikh, H.R., Simoncelli, E.P.: Image quality
  assessment: from error visibility to structural similarity. Transactions on
  Image Processing  \textbf{13}(4),  600--612 (2004)

\bibitem{wollmann2021}
Wollmann, T., Rohr, K.: {Deep Consensus Network: Aggregating predictions to
  improve object detection in microscopy images}. Medical Image Analysis
  \textbf{70},  102019 (2021)

\bibitem{xiao2018simple}
Xiao, B., Wu, H., Wei, Y.: Simple baselines for human pose estimation and
  tracking. In: European conference on computer vision (ECCV). pp. 466--481
  (2018)

\bibitem{xie2018microscopy}
Xie, W., Noble, J.A., Zisserman, A.: Microscopy cell counting and detection
  with fully convolutional regression networks. Computer methods in
  biomechanics and biomedical engineering: Imaging \& Visualization
  \textbf{6}(3),  283--292 (2018)

\bibitem{yang2020circlenet}
Yang, H., Deng, R., Lu, Y., Zhu, Z., Chen, Y., Roland, J.T., Lu, L., Landman,
  B.A., Fogo, A.B., Huo, Y.: {CircleNet: Anchor-Free Glomerulus Detection with
  Circle Representation}. In: International Conference on Medical Image
  Computing and Computer-Assisted Intervention (MICCAI). pp. 35--44. Springer
  (2020)

\bibitem{zhou2019objects}
Zhou, X., Wang, D., Kr{\"a}henb{\"u}hl, P.: Objects as points. arXiv preprint
  arXiv:1904.07850  (2019)

\end{thebibliography}
\end{document}